\documentstyle [draft,mathtools]{mn}
\title[Dynamical and radiative simulations of $\gamma$-ray jets in microquasars]
      {Dynamical and radiative simulations of $\gamma$-ray jets in microquasars}
\author[T. Smponias \& T. \ S. Kosmas]
      {T. Smponias$^{1}$ \& T. S. Kosmas$^{1}$\thanks{E-mail:hkosmas@uoi.gr}\\
      $^{1}$Physics Department, University of Ioannina, GR-45110 Ioannina,
      Greece}

\date{Received 1 January 0000; in original form 0000 February 1}
\pagerange{\pageref{firstpage}--\pageref{lastpage}}

\def\LaTeX{L\kern-.36em\raise.3ex\hbox{a}\kern-.15em  T\kern-.1667em\lower.7ex\hbox{E}\kern-.125emX}

\let\leqslant=\leq 

\newcommand{\aap}{Astron. \& Astrophys.}

\newcommand{\apj}{Ap. J.}
\newcommand{\mnras}{MNRAS}

\newcommand{\nat}{Nature}
\newcommand{\apjl}{Ap. J. Lett.}
\newcommand{\araa}{Ann. Rev. Astron. Astrophys}

\newcommand{\prd}{Phys. Rev. D}

\newcommand{\apjs}{Ap. J. Supp.}
\newcommand{\apss}{Astrophys. \& Space Science}

\newcommand{\nar}{New Astron. Rev.}
	
\newcommand{\baas}{Bulletin of the American Astronomical Society}

\begin{document}

\maketitle

\begin{abstract}
The emission of $\gamma$-rays in jets emanating from the vicinity of collapsed stellar remnants, in binary systems known as microquasars, is investigated using a three dimensional relativistic hydrocode (PLUTO), in combination with two in-house radiative transfer codes. Even though a great number of stellar systems may be addressed by such models, we restrict ourselves to the concrete example of the SS433 X-ray binary, the only microquasar with a definite hadronic content in its jets, as verified from spectral line observations. A variety of system configurations have been examined, by employing a hadron-based emission mechanism.
The dependence of the $\gamma$-ray emissions on certain dynamical source properties, such as the hydrodynamic parameters of the mass-flow density, the gas-pressure and the temperature of the ejected matter, is simulated. Radiative properties, especially the assumed high energy proton population inside the jet plasma, and its effect on the calculated emission, are also examined.
Two sets of initial conditions of the chosen microquasar are employed, in order to cover different scenarios pertaining to the system under consideration.
\end{abstract}

\begin{keywords}
ISM: jets and outflows--stars: winds, outflows--stars: --gamma-rays: theory--stars: flare-- radiation mechanisms: general.
\end{keywords}

\section{Introduction}
\label{HDmodelling}

In the past two decades, the emissions of $\gamma$-rays, particles and neutrinos within jets in microquasars have been extensively investigated. The mechanisms suggested for the underlying emission processes involved either photon-hadron interactions (Levinson \& Waxman 2001, Distefano et al. 2002), or hadron-hadron interactions (Romero et al. 2003, Christiansen, Orellana \& Romero 2006, Reynoso, Romero \& Christiansen 2008, Romero \& Vila 2008, Vila \& Romero 2010, Vila, Romero \& Casco 2012). Similar mechanisms for $\gamma$-ray absorption have been suggested (Reynoso et al. 2008, Cerutti et al. 2011). From an observational point of view, $\gamma$-rays from microquasars may be seen with both (terrestrial) Cherenkov telescopes, such as HESS (Hinton 2004) and new HESS II (HESS collaboration, 2012), MAGIC (Baixeras et al. 2004) and VERITAS (Weekes et al. 2002), and the upcoming CTA (Actis et al. 2011), as well as with orbital telescopes (for an earlier medium energy observation see Geldzahler et al. 1984), like ESA sattelite \emph{INTEGRAL} and NASA orbital telescope \emph{Fermi} (e.g. Abdo et al. 2009). For further information, the reader is referred to Orellana et al. (2007a), Hayashi et al. (2009), Saito et al. (2009). Concerning neutrino emission, although detection of such particles from microquasars has not been clearly achieved to date, estimates for their intensity have been derived (Aiello et al. 2007). This may serve as a guide for future attempts to detect, but also to model high energy neutrino and $\gamma$-ray emissions from galactic microquasars (e.g. Mirabel, 2006, Reynoso et al. 2008). In the future, relevant data may be collected even from extra-galactic microquasars (Joseph, Maccarone \& Fender 2011, Muxlow et al. 2011), as $\gamma$-ray and neutrino observation/detection sensitivities improve. The $\gamma$-ray emission mechanisms might, in fact, be even stronger, since magnetic field effects in the jet may act towards considerably reducing total emission within the jet, by enhancing processes that limit the high energy proton population (an example is the synchrotron radio emission, Reynoso \& Romero 2008). 

In general, the jets of compact microquasars may be considered as fluid flow emanating from a central source at the jet origin (Mirabel \& Rodriguez 1999), an approach often used for describing astrophysical jets. Such a well known system is the SS433 microquasar (Begelman et al. 1980; Margon 1984; Fabrika 2004), an X-ray binary comprising a supergiant star and a collapsed stellar remnant. This system emits relativistic jets which have been studied in various wavelength bands. So far, it is the only microquasar (MQ) observed with a definite hadronic content in its jets, as verified from observations of spectral lines (Begelman et al. 1980; Margon 1984; Fabrika 2004). Further examples of systems that can be modelled in a similar way, for $\gamma$-ray and high energy neutrino emission, are the microquasar LSI +61 303 (Christiansen et al. 2006) that, compared to SS433, has a faster jet flow, but weaker in terms of mass output rate, and also the Cyg X-3 system (Tavani et al. 2009, Abdo et al. 2009). The galactic microquasar GX339-4 (NEMO experiment, Aiello et al. 2007) is another possible site for high energy neutrino production within its jets.

In a relativistic treatment of the jets, a variety of mechanisms of energy loss through relevant hadronic processes may be adopted. On a macroscopic scale, the jet matter may be assumed to behave as a fluid (fluid approximation), due to the magnetic field creating 'mesoscopic' coupling, leading to collective dynamic behavior (Shu 1991, Ferrari 1998, De Young 2002). At a smaller scale, however, the kinetic description of the plasma becomes necessary for the treatment of effects at shock acceleration sites (Rieger, Bosch-Ramon \& Duffy 2007). One may consider the scenario where proton-proton (pp) collisions (between a non-thermal high energy proton and a slower, bulk flow one), represent the dominant cooling process of the high energy proton population, producing the respective $\gamma$-rays and neutrinos in the binary SS433 system (Rieger, Bosch-Ramon \& Duffy 2007, Bosch-Ramon et al. 2006, Reynoso et al. 2008). The acceleration of thermal protons takes place only for those having energy above a given threshold (Reynoso et al. 2008) that allows diffusive first-order shock acceleration to occur (Rieger et al. 2007). Assuming a Maxwellian energy distribution for the 'slow' protons (those that constitute the jet flow), only the fastest shall be eligible for diffusive shock acceleration, therefore, just a tiny portion of the whole bulk proton jet-flow is transferred to the non-thermal proton population. Consequently, the non-thermal protons represent a small fraction of the total jet proton density (Bosch-Ramon et al. 2006, Reynoso et al. 2008).

Another significant mechanism for the generation of high-energy $\gamma$-rays in microquasars has been suggested (Romero et al. 2003), which is based on hadronic interactions occurring within the jet-wind interaction zone. According to this mechanism, the $\gamma$-ray emission arises from the decay of relativistic neutral pions
\begin{equation}
\pi^0 \to \gamma + \gamma .
\end{equation}
Pions are created via inelastic collisions of jet protons, ejected from the compact object, and ions in the stellar wind. This process may also occur in the vicinity of the extended disk of the binary system (Romero et al. 2003), even though, recently, it has been argued that such an emission scenario is rather unlikely to occur in SS433 (Reynoso et al. 2008). The simulations performed in the present work, for $\gamma$-ray emission from the SS433 MQ, are motivated by the prospect of improved new data. For $\gamma$-rays of lower energy, observations with \emph{Fermi} and \emph{INTEGRAL} (ESA satellite) may offer new data, while higher energy $\gamma$-rays, generally above the threshold of 30 \rm{GeV}, can be seen with ground based Cherenkov telescopes, such as HESS, MAGIC and the CTA, that may allow further insight into the properties of the SS433 system.
 
In some recent modelling attempts for $\gamma$-rays and neutrinos from microquasars, a conical or, in general, a fixed geometry jet, is perceived. The properties of the jet vary along its length, depending on the position and the opening angle (for a more detailed description of the geometry of a similar model, see Hjellming and Johnston, 1988). Radiative transfer calculations are then performed at every point in the jet, providing (for a range of wavelengths or energies at every location) the relevant emission and absorption coefficients. Subsequently, a line-of-sight integration process leads to a synthetic image of the jet's $\gamma$-ray emission, at a given energy observing window. Total jet emission at that frequency can then be obtained by summing over the pixels of the synthetic image. Simulations like the above often assume a steady state jet as the geometrical basis for further radiative calculations leading to output of observable and/or detectable quantities. A natural next step is therefore, to use a more realistic underlying jet model as a basis for the emission/absorption computations. An advantageous jet model may be a hydrodynamic one, offering additional information for the structure of the jet and its surrounding environment, as opposed to a pre-set geometry jet. 

In the present paper, we address emission of $\gamma$-rays in the energy range from sub-\rm{GeV} up to more than $10^{2}$ \rm{TeV}, modelled with a combination of a three dimensional relativistic hydrocode, PLUTO (Mignone et al. 2007), and two in house radiative transfer codes.  The calculation of the emission and absorption coefficients (Reynoso et al. 2008), is based on the use of the Monte Carlo simulations of elementary particle reactions (Kelner et al. 2006) that lead to $\gamma$-ray production in microquasar jets. Those simulations provide relatively concise expressions for the emission and absorption coefficients, at various $\gamma$-ray frequencies. Instead of a steady-state model, the formalism of Kelner et al. (2006) is now applied to a hydrodynamical jet, based on the properties of the emitting jet body. More specifically, we use the line-of-sight (LOS) code to obtain $\gamma$-ray emission/absorption coefficients at every jet location, based in part on hydrodynamic variables supplied by the hydrocode. The combined result includes the ‘effect’ of the high energy proton distribution on the hydrodynamical model jet, as viewed through the integration. 

The paper is organized as follows. At first (Sections \ref{formalism}, \ref{hydrodynamicaljet}, \ref{radtransfer}) the main formalism is presented and the features of our method are summarized. Then, in Section \ref{results}, we present and discuss the results obtained, (i) with the steady-state model, and (ii) with the 3-D relativistic hydrocode. The output of the latter is afterwards inserted into the in-house radiative code and the 3D data are subject to LOS integration. During this integration, the LOS code communicates with a program which provides the emission coefficients at every model point. We finally summarize of the main conclusions extracted from our calculations. 

\section{ Description of the formalism }
\label{formalism}

The starting point of the formalism is the general radiative transfer equation describing the emission, scattering and absorption of particles, including neutrinos, and electromagnetic radiation, given by the differential equation (see Chandrasekhar 1960, Sharkov 2004) 

\begin{equation}
[\frac{\partial}{\partial t} +\hat{\Omega}\cdot \nabla +(k_{\rm{\nu,s}}+k_{\rm{\nu,a}})]I_{\rm{\nu}}=j_{\rm{\nu}}+k_{\rm{\nu,s}} \frac{1}{4\pi } \int\limits_{\Omega} I_{\rm{\nu}} d\Omega
\label{radtransfergeneral}
\end{equation}
where $j_{\rm{\nu}}$ and $k_{\rm{\nu}}$ denote the local emission and absorption coefficients, respectively. On the l.h.s., of the latter equation, the first term in the brackets represents the time rate of change of the radiative transfer ($u << c$). In the second term the divergence (spatial variation) of the radiation field is included. $\Omega$ represents the solid angle subtended by the radiation beam considered. The third term holds the intensity loss due to absorption ($k_{\rm{\nu, \alpha}}$) and/or scattering ($k_{\rm{\nu, s}}$). On the r.h.s., the second term takes into account the 'secondary' emissions due to scattering.

In the case of time independent radiative transfer, and absence of sideways scattering, when moving along the direction of a one-dimensional LOS, Eq. \ref{radtransfergeneral} reduces to 
\begin{equation}
[\hat{\Omega}\cdot \hat{\eta} \frac{d}{dl} +k_{\rm{\nu,a}}]I_{\rm{\nu}}=j_{\rm{\nu}}.
\label{radtransfersimple}
\end{equation}
where $\hat{\eta}$ is a unit vector in the direction of the LOS.
 Along a LOS that subtends a solid angle $\Delta \Omega$, equation \ref{radtransfersimple} becomes
\begin{equation}
 \Omega_{LOS} \cdot \frac{dI_{\rm{\nu}}}{dl}= j_{\rm{\nu}}- k_{\rm{\nu}} I_{\rm{\nu}}
 \label{radtransferlos}
\end{equation}
 The two coefficients, $j_{\rm{\nu}}$ and $k_{\rm{\nu}}$, are, in general, dependent on the properties of the jet matter. Consequently, in order to calculate the emission from a given  microquasar jet, one also needs information for the geometrical and dynamical conditions of the system in question.

\subsection{Model jet emission and dynamics}
\label{modeljetdynamics}

For both steady state jet and time dependent jet models, we assume that the time scale of the radiative transfer is much smaller than the system's dynamical one. Consequently, the time independent form  of the radiative transfer equation (\ref{radtransfersimple}) can be reliably applied. As a first approximation for the SS433 system the jet speed, $u_{\rm{jet}}$, is taken to be significantly smaller than the speed of light, $\beta=u_{\rm{jet}}/c \simeq$ 0.26, and we use our (non-relativistic) radiative transfer code (Smponias \& Kosmas 2011). Based on the assumption $u_{\rm{max}} << c$, where $u_{\rm{max}}$ is the maximum speed anywhere in the system, which in our case is essentially the jet speed. The time-independent radiative transfer  equation (\ref{radtransferlos}) is then applicable. The relatively smooth flow of the model jet, as it appears in the simulations, suggests that the jet does not change significantly during the  light crossing time of the jet perpendicular to its axis. Matter outside the jet is clearly sub-relativistic, therefore it does not constitute a problem in terms of relativistic imaging.

\subsubsection{Characteristics of the steady state model jet}\label{steadystatemodel}

\begin{table*}
\begin{tabular}{||c||r|c|c|}
\hline\hline
\multicolumn{1}{|c|}{ Parameter/Scenario} & 
\multicolumn{1}{|c|}{A (run1)} &
\multicolumn{1}{|c|}{B (run2)} & 
\multicolumn{1}{|c||}{ Comments} \\
\hline
\multicolumn{1}{|c|}{cell size ($\times 10 ^{10} \rm{cm}$)} & 0.25  & 0.40 &  PLUTO's computational cell \\
\multicolumn{1}{|c|}{ $\rho_{\rm{jet}}$  (\rm{cm}$^{-3}$)}  & $1.0\times 10^{11}$ & $1.0 \times 10^{14}$ & initial jet matter density   \\
\multicolumn{1}{|c|}{ $\rho_{\rm{sw}}$ (\rm{cm}$^{-3}$)}  & $1.0\times 10^{12}$  & $1.0 \times 10^{12}$ & stellar wind density \\
\multicolumn{1}{|c|}{ $\rho_{\rm{adw}}$ ($\rm{cm}^{-3}$)}  & $1.0\times 10^{12}$  &  $1.0 \times 10^{13}$ & accretion disk wind density  \\
\multicolumn{1}{|c|}{$t^{\rm{max}}_{\rm{run}}$ (\rm{s})}  & $1.5 \times 10^{3}$  & $1.5 \times 10^{3}$  & model execution time \\

\multicolumn{1}{|c|}{Method}  & P. L. & P. L. & Piecewise Linear \\
\multicolumn{1}{|c|}{Integrator}  & Ch. Tr. &  Ch. Tr. & Characteristic Tracing \\
\multicolumn{1}{|c|}{ EOS}  & Ideal &  Ideal & Equation of state\\
\hline
\multicolumn{1}{|c|}{ $n$}  & 0.1005 & 0.1005 & $E_{\gamma}$=100 
\rm{GeV} normalisation \\
\multicolumn{1}{|c|}{ BinSep (\rm{cm})}  & $4.0 \times 10^{12}$  & $4.0 \times 10^{12}$  & Binary star separation \\
\multicolumn{1}{|c|}{ M$_{\rm{BH}}$/$M_{\sun}$}  & 3-10 & 3-10 & Mass range of collapsed star\\
\multicolumn{1}{|c|}{ M$_{\star}$/$M_{\sun}$}  & 10-30 & 10-30 & Mass range of Main Seq. star \\
\multicolumn{1}{|c|}{ $\beta = v_{0}/c$ } & 0.26 & 0.26 & Initial jet speed \\
\multicolumn{1}{|c|}{ $L^{\rm{p}}_{\rm{k}}$} & $ 10^{37}$ & $ 10^{40}$ & Jet kinetic luminocity \\
\multicolumn{1}{|c|}{ grid resolution}  & 480*720*480  & 300*500*300 & PLUTO grid resolution (xyz)  \\
\hline \hline
\end{tabular}
\caption{Values of various physical and model parameters (first column) pertaining to the $\gamma$-ray emission from the jet, for two simulation runs, obtained with the piecewise linear method. Run 2 has a slightly longer grid, while in Run 1 all the system's densities are decreased by a few orders of magnitude, in order to account for a lower jet mass flow rate (jet's kinetic luminocity).$n$ stands for normalization between the two different methods of calculation of the $\gamma$-ray emission coefficient, one valid for energies above $E_{\gamma}$=100 \rm{GeV} and the other below (Eqs \ref{O1}, \ref{O2}).}
\label{Table1}
\end{table*}

Many authors have considered the high-energy (up to 100 TeV) $\gamma$-ray emission from relativistic hadrons in microquasars, to emerge from the interaction of relativistic protons in the jets with target protons of the companion's stellar wind, usually employing a steady-state jet model (Romero et al. 2003; Christiansen, Orellana \& Romero 2006; Orellana et al. 2007b). In the case of the binary SS433 system, however, which comprises two oppositely directed, precessing jets with hadronic content (the stellar wind is estimated to be rather weak, Reynoso et al. 2008), therefore, the $\gamma$-rays and neutrinos may be assumed to be created mainly from proton-proton interactions between relativistic (fast) and thermal (cold) protons within the jets themselves, without excluding other production mechanisms, though. 

In the present work we adopt the usual assumption that a small but very energetic proton population, $n_{\rm{fp}}$, created at randomly oriented shock fronts in the jet, interacts with the bulk flow protons of the jet. This also allows to explore the jet-wind interaction  mechanism, which is now governed by the hydrodynamical conditions pertaining to such interactions. The presumed population of high energy protons is produced from the bulk flow of jet protons, through first order Fermi acceleration taking place at shocks within the jet (Rieger et al. 2007, Blandford \& Eilek 1987, Reynoso et al. 2008). These shocks are approximately taken  as homogeneously distributed throughout the jet, effectively linking the jet matter density to the density of the shocks.

The internal shocks do convert some of the bulk kinetic energy of cold protons, K, to the energy of the multi-directional motion of the non-thermal protons, $E_{\rm{p}}$. In fact, for the rate $t_{\rm{accel}}^{-1}$, at which some among the thermal protons populate their high energy counterparts' distribution, we adopt the relation (Begelman et al. 1990; Reynoso et al. 2008)

\begin{equation}
t_{\rm{accel}}^{-1}\simeq  \beta^{2} \frac{ceB}{E_{\rm{p}}},
\label{taccel}
\end{equation}
where $\beta \sim \Delta u_{\rm{}}/c$, $B$ stands for the magnetic field and $e$ is the proton charge. $\Delta u$ is the difference between the (often relativistic) upstream velocity and the non-relativistic downstream velocity of a shock in the jet. The upstream one is the jet bulk speed, whereas the downstream one is of order of the shock speed (Rieger et al. 2006). Therefore, $\Delta u$ can be taken as equal to the upstream velocity itself. Equation \ref{taccel} gives the production rate of non-thermal protons at every point in the jet. However, the production of $\gamma$-rays from these non-thermal protons occurs as a next stage. Consequently, equation (\ref{taccel}) provides an upper limit for the production rate of $\gamma$-rays, through several cooling processes (Reynoso et al. 2008, fig. 2). The rate of the latter cannot of course exceed, order-of-magnitude-wise, the acceleration rate of non-thermal protons at a given jet location (the 'location', is determined by a hydrodynamical grid cell, which, microscopically, is still very large, of order 10$^{10}$ cm. Therefore, average values for particle populations apply to each cell). 

Employment of $\beta^{2}$ in Eq. \ref{taccel}, cuts off $\gamma$-ray emission from slow moving matter, since the  acceleration efficiency (in randomly oriented internal shocks) is less efficient in slow matter. Nevertheless, one must bear in mind that, the proton acceleration rate cannot affect the $\gamma$-ray emission rate, unless $\beta$ drops below some value. For a steady-state jet, one may conclude (figure 2 of Reynoso et al. 2008) that, for lower energies the emission rate is several orders of magnitude below the acceleration rate. However, for higher energies, a smaller local velocity may force the acceleration rate to drop below the emission rate. Towards this purpose, we introduce a factor $w$, which can be set equal to either the local jet density, $w=\rho$, or to the density times the square of the relativistic velocity, $w=\rho \beta^{2}$. The velocity filter enhances the features of the jet where the matter moves faster locally. Different imaging methods can be applied, by selecting a different expression for $w$ (see below, Section \ref{runs}). In this paper, we adopt the approximation in which, when the non-thermal proton production rate goes below the $pp$ energy-loss rate (figure 2 of Reynoso et al. 2008), the energy loss rate becomes equal to the non-thermal protons' production rate. Reynoso et al. (2008) address a number of energy loss processes for energetic protons in the jet. As long as the local velocity of the jet matter u$_{jet}$ is relativistic, the primary process for energy loss is the pp-collision, that may be considered as the main $\gamma$-ray emission mechanism. For slower protons, other energy loss mechanisms become important, even though the overall emission decreases, so that the energetic proton production limit may be taken as a ceiling for energy loss. 

It should be noted that our model jet may incorporate additional processes for $\rm{gamma}$-ray production, such as leptonic secondary emission (Orellana et al. 2007b), in which charged pion-decay produces relativistic leptons that become trapped in the jet by the magnetic field and subsequently create $\rm{\gamma}$-rays, via processes similar to those considered in primary leptonic models. Furthermore, neutral-pion decay $\rm{\gamma}$-rays are absorbed by soft X-rays originating from the donor star (Cerutti et al. 2011), which leads to electron-positron pairs that in turn cause inverse Compton scattering of soft X-rays photons, eventually producing $\rm{\gamma}$-rays of different energies than their progenitors. 

As was mentioned earlier, the ejected jet particles are mostly slow (thermal) protons, of density $n_{\rm{sp}}$, or simply $n_{\rm{p}}$, but there is a small portion of non-thermal (high energy) protons, of density $n_{\rm{fp}}$, swirling among them (Bosch-Ramon et al. 2006). The form of energy distribution of the non-thermal protons, in the jet's frame of reference, is: 
\begin{equation}
n_{\rm{fp}}(E^{\prime}_{\rm{fp}}) = K_{0}(E_{\rm{fp}}^{\prime})^{-\alpha}
\label{Eq-powerlaw}
\end{equation}
(Reynoso et al. 2008, Christiansen et al. 2006) where $K_0 $ is a normalization constant that can be determined as described below. The parameter $\alpha$, which denotes a power-law type distribution, takes the value of $\alpha = 2$ (Rieger et al. 2007). For a steady-state jet expanding laterally with an opening angle diminishing along its axis (Hjellming and Johnston 1988, Romero et al. 2003), we have:
\begin{equation}
 \left( \frac{z_{\rm{0}}}{z_{\rm{j}}} \right)^{2 \ell} = \left( \frac{R_{\rm{0}}}{R_{\rm{j}}} \right)^2=\frac{A_{\rm{0}}}{A_{\rm{j}}}
\label{Eq-conicmodel}
\end{equation}
In the latter equation, $z_{\rm{0}}$ is the position along the jet axis, of a reference point,  $R_{\rm{0}}$ is the jet radius at $z_{\rm{0}}$, and $A_{\rm{0}}=\pi R_{\rm{0}}^{2}$ is the jet cross-section at $z_{\rm{0}}$. $z_{\rm{j}},R_{\rm{j}},A_{\rm{j}}$ are the above quantities at any jet point. The power $\ell$ determines the rate of the jet's sidereal expansion. Setting $\ell$=0 produces a cylindrical jet (no expansion), whereas $\ell$=1 produces a conical jet (unimpeded sidereal expansion, as in Christiansen et al. 2006 and in Reynoso et al. 2008). The reader is referred to Hjellming and Johnston (1988) for a model jet with similar geometry.

As implied from equation (\ref{taccel}), the acceleration rate $r=t_{\rm{acc}}^{-1}$ of non-thermal protons is dependent on the magnetic field B. Since our simulations are non-magnetised, we could assume, as a rule of thumb, that B is inversely proportional to z, i.e. $B\simeq z^{-1}$ (Rieger et al. 2007, see also Bosch-Ramon et al. 2006). As far as the hydrodynamical model of the next section is concerned, even though a more accurate dependence of $B$ on the location along the jet axis could be adopted, the above dependence is still considered to be adequate. In this way, $t_{\rm{acc}}^{-1}$, does not change more than 1-2 orders of magnitude over the range of spatial scales of the model jet environment. Thus, the acceleration mechanism can balance the energy losses of the non-thermal proton population. The acceleration rate r (at a given jet proton density), is proportional to the square of the local velocity of the fluid and inversely proportional to the density, which in turn decreases along the jet (see equation \ref{zn} below). On the other hand, the rate at which these particles react, and thus lose energy, producing pions and $\gamma$-rays, is subject to a number of processes (Reynoso et al. 2008).

\subsubsection{Calculation of fluxes in the steady-state model jet}

We denote, in the frame of observation, the flux densities of the two proton populations, non-thermal and thermal, as $J_{\rm{fp}}$ and $J_{\rm{sp}}$, respectively, and in the jet's frame $J^\prime_{\rm{fp}}$ and $J^{\prime}_{\rm{sp}}$. At a given point in the moving (jet) frame, the energy spectrum of the non-thermal protons is then described by equation (\ref{Eq-powerlaw}), while the spatial density is given by the relation
\begin{equation}
w n_{\rm{fp}}(E') dE' =w K_{\rm{0}}(E')^{-\alpha} dE'
\label{zn}
\end{equation}
where (see also Section \ref{steadystatemodel})
\begin{equation}
w = n_{\rm{sp}} \beta^{2}
\label{wdefinition}
\end{equation}
and $\beta \equiv u/c$. The magnitude of the local velocity vector, $\vert {\bf u} \vert$, is $ u=\sqrt{u_{\rm{x}}^2 + u_{\rm{y}}^2 + u_{\rm{z}}^2}$. The variation with $z_{\rm{j}}$ of the flux of non-thermal protons, $J^{\prime}_{\rm{fp}}$, for a jet with an opening angle diminishing along its axis reads (Romero et al. 2003):
\begin{equation}
J^\prime_{\rm{fp}} = \frac{c}{4\pi} K_0 \left( \frac{z_{\rm{0}}}{z_{\rm{j}}} \right)^{2 \ell}
\left( E^\prime_p \right)^{-\alpha} \beta^{2}
\label{Eq-Jfp}
\end{equation}
where we have assumed that $J'_{\rm{fp}}$ is inversely proportional to the area of the jet cross section A. For a steady-state jet, the velocity u is constant and could be incorporated in the respective normalization constant $K_{0}$ (this is not possible for a time-dependent hydrodynamical model). It should also be noted that both $J'_{\rm{fp}}$ and $J'_{\rm{sp}}$ vary proportionally to $( z_0 / z_{\rm{j}} )^{2 \ell}$ in the steady state model jet, but $J'_{\rm{fp}}$ bears an additional dependence on $K_{0}({E'}_{p})^{-\alpha}$, representing the power-law energy-dependence of the emitted flux of the non-thermal protons. At $z_j = z_0$ Eq. \ref{Eq-Jfp}
takes the form
\begin{equation}
J^{\prime}_{\rm{fp}} = \frac{c}{4\pi} K_{\rm{0}} \left( E^{\prime}_{p} \right)^{-\alpha} \beta^{2} \, .
\label{Eq-Jfp2}
\end{equation}
In order to calculate the normalization constant $K_{0}$, we need to fix the model jet volume so as to match the observed $\gamma$-ray emitted power, at a given frequency interval, to the corresponding power obtained from the model. Obviously, $J_{\rm{sp}} = n_{\rm{sp}} u_{\rm{sp}} = I/A$, where $u_{\rm{sp}}$ stands for the thermal (bulk flow) proton velocity, $I$ is the (total) mass current ($I$ and u$_{\rm{sp}}$ are considered constants throughout the jet). Using equation (\ref{Eq-conicmodel}), we can write for a position $z$ on the z-axis,  
\begin{equation}
 \left( \frac{ z_{\rm 0} }{z} \right)^{2 \ell} = \left( \frac{R_{\rm{0}}}{R} \right)^2 = \frac{A_{\rm{0}}}{A}=\frac{J}{J_{0}} =\frac{ n_{\rm{sp}} } { n_{\rm 0} }
\label{cone-crosec3}
\end{equation}
%
which relates the jet geometry to the properties of the jet matter.

\section{A new dynamical and radiative jet model}
\label{hydrodynamicaljet}

In this section, on the basis of the steady-state jet model, we will formulate our dynamical and radiative $\gamma$-ray emission model. We start from equation (\ref{wdefinition}) which implies that $\gamma$-ray emission from fast-moving matter should be enhanced compared to the steady-state, while emission from slower matter should be lower. The reasoning behind the above is that, in equation (\ref{taccel}) $t_{\rm{acc}}$, being proportional to $\beta^{2}$, creates non-thermal proton density in the jet, which in turn allows for pp-collisions to occur leading to $\gamma$-rays. $\beta^{2}$, then, acts as a ceiling for $\gamma$-ray emission.

 We note that, in the general case, the local velocity $u$ is not necessarily parallel to the y-axis. For example, in case of local turbulence in the vicinity of a computational cell, the velocity of in-cell matter may point almost anywhere. Meanwhile, emission is caused by jet matter moving towards any direction (emission mechanism is based on randomly oriented turbulent shocks). Hence, the emission is taken to be multi-directional and no secondary emissions from scattering are considered (see Eq. \ref{radtransfergeneral}). In addition, more shocks occur wherever the jet matter moves faster. The largely turbulent nature of the jet flow (as it also appears in our simulations, especially at higher resolutions), offers the locale for shocks to exist in. A strong dependence on local velocity, is achieved through dependence on $\beta^{2}$. We arrive at that expression (Eq. \ref{wdefinition}) by the use of non-thermal proton acceleration rate as a ceiling for energy loss from the non-thermal proton distribution. As the acceleration rate is proportional to $\beta^{2}$, the energy loss rate is also set proportional to $\beta^{2}$ and then the emission due to that is proportional to $J = \rho \beta^{2}$. The above results shall be used with the radiative calculations later on (see Section \ref{radcoderesults}). The spatial density of non-thermal protons, $n_{\rm{fp}}$, is given by equation \ref{zn}.

We now replace the quantity $(z_0/z_{\rm{j}})$, in Eq. \ref{Eq-Jfp}, by a dependence on hydrodynamic quantities, namely the density and the velocity which are calculated everywhere by the PLUTO code. The quantity $z$ of the static model is replaced by the quantity $y$ of the hydrodynamic model, since in the latter the jet is taken to propagate along the $y$ direction. By combining equations (\ref{Eq-Jfp}) and (\ref{cone-crosec3}), for the hydrodynamical jet, the non-thermal proton current density, $J^{\prime}_{\rm{fp}}(E^{\prime})$ takes the form
\begin{equation}
J^{\prime}_{\rm{fp}}=\frac{c}{4\pi}K_{0} \frac{n_{\rm{j}}}{n_{\rm{0}}} (E')^{-\alpha} \beta^{2} 
\label{jprimefp1}
\end{equation}
where $n_{\rm{j}}$ is the local density of the (thermal) bulk jet protons.
A new normalization constant, $K_{1}$, may then be then defined,
\begin{equation}
K_{1}=\frac{K_{\rm{0}}}{n_{\rm{0}}} \, ,
\label{Jfp-2}
\end{equation}
$K_{1}$ shall be adjusted at the end, in order to match the observed/upper limit for the total jet emission to the observed one (Saito et al. 2009). The density of the jet, $\rho_{\rm{jet}}$, is set hydrodynamically, according to estimates taken from the literature (Begelman et al 1980, Fabrika 2004).
Also, from Reynoso et al. (2008) we have
\begin{equation}
K_{0}=\frac{4 q_{\rm{rel}} L_{\rm{k}}}{cR_{\rm{0}}^{2}ln(E'_{\rm{max}}/E'_{\rm{min}})}
\label{args08equation5} \, ,
\end{equation}
where $L_{\rm{k}}$ is the kinetic luminocity of the jet, and $E'_{\rm{min}}$ and $E'_{\rm{max}}$ are the minimum and maximum proton energies. Furthermore $K_{1}$ may be expressed as
\begin{equation}
K_{1}=\frac{4 q_{\rm{rel}}L_{\rm{k}}}{\beta^{2}_{0} n_{\rm{sp0}}c R_{0}^{2}ln(E'_{\rm{max}}/E'_{\rm{min}}) }
\label{args08equation5adapted}
\end{equation}
where $n_{\rm{sp0}}$ is a reference value for the proton number density, taken at the point $z_{0}$. The factor q$_{\rm{rel}}$ in Eqs. \ref{args08equation5} and \ref{args08equation5adapted}, stands for the fraction of the energy carried by the non-thermal protons. A value of $q_{\rm{rel}}$=0.0001 is initially adapted. Nevertheless, this parameter remains unconstrained this way (Saito et al. 2009) and therefore needs to be determined by employing an external argument for $K_{1}$. 

\subsubsection{Laboratory frame description}

\begin{figure} 
\vspace*{4.7cm}
\centering
 {\includegraphics{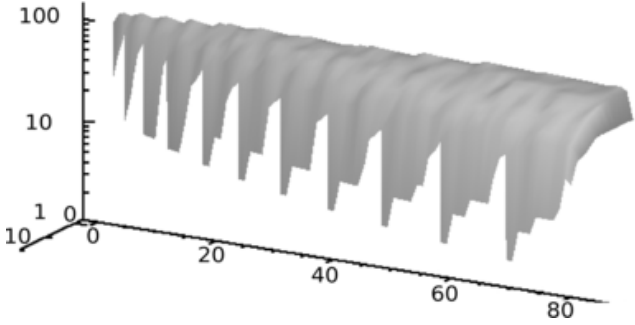}}
\caption{ A 3D LOS synthetic image of the steady state jet, formed using the geometry of the Hjellming \& Johnston (1988) model (verical $z$ axis in arbitrary units, $x$ and $y$ axes are in 5$\times$10$^{10}$ cm Cartesian coordinates), and then running the LOS code of Smponias \& Kosmas (2011) through it (a resolution of $60 \times 100 \times 60$ is employed).}
\label{losimagehj88}
\end{figure}

The transformation of equations (\ref{Eq-Jfp}) and (\ref{jprimefp1}) to the observer's frame (using equations \ref{cone-crosec3} and \ref{Jfp-2}), reads
\begin{equation}
J_{\rm{fp}} (E_p, Z_j, t) = \beta^{2} \left( \frac{z_{\rm{0}}} {z_{\rm{j}}} \right)^{2 \ell} \frac{c}{4\pi}K_{0} S = \frac{c}{4 \pi} K_{1} n_{\rm{sp}} \beta^{2} S 
\label{jfpfull}
\end{equation}
where $S$ is a function of stationary frame energy $E_{\rm{p}}$, given by (Reynoso et al. 2011)
\begin{equation}
S = \Gamma\left( E_{\rm{p}} - \beta\cos{i_{\rm{j}}} \sqrt{E_{\rm{p}}^2 - m_{\rm{p}}^2c^4} \right) 
\end{equation}
In the latter equation, $i_{\rm{j}} (t)$ denotes the angle between the jet axis and the LOS. The jet bulk Lorentz factor $\Gamma$ is 
\begin{equation}
\Gamma=\frac{1}{\sqrt{1-\beta^{2}}} \, ,
\end{equation}
$S$ provides an expression for our own frame's J$_{\rm{fp}}$, that depends on the $\gamma$-ray energy, $E_{\rm{\gamma}}$ as measured in the laboratory frame. Therefore, from now on we can work with laboratory frame quantities only, which are also the jet model quantities. 

\section{Development of the radiative transfer formalism}
\label{radtransfer}

In the present work, we consider propagation of $\gamma$-rays without scattering along a one-dimensional line-of-sight. The basic equation needed is the one-dimensional version of equation (\ref{radtransfergeneral}), without the scattering terms. Assuming that $\Omega_{\rm{LOS}}$ is the same for every LOS, in the $\gamma$-ray notation Eq. \ref{radtransfergeneral} is written as
\begin{equation}
\frac{dI_{\rm{\gamma}}}{dl}=\epsilon_{\rm{\gamma}}-I_{\rm{\gamma}}\kappa_{\rm{\gamma}}
\label{1dradtransfer}
\end{equation}
We shall first calculate the emission from a small jet element cell of volume $dV$, by defining the quantity 
\begin{equation}
dI_{\rm{\gamma}}=J_{\rm{\gamma}} dV=\rho \frac{dN_{\rm{\gamma}}}{dE_{\rm{\gamma}}} dV= \frac{dN_{\rm{\gamma}}}{dE_{\rm{\gamma}}} dm_{\rm{cell}}
\label{21blackjack}
\end{equation}
to represent (apart from a normalization constant) the intensity emanating from a jet cell of volume $dV$ at a given $\gamma$-ray energy $E_{\rm{\gamma}}$. $\rho$ is the hydro density of the cell ($\rho$=$m_{\rm{p}} n_{\rm{sp}}$) and $dN_{\rm{\gamma}}/dE_{\rm{\gamma}}$ stands for the differential emissivity of the cell at $E_{\rm{\gamma}}$. As we have discussed in Section \ref{hydrodynamicaljet}, an alternative version of the above quantity is 
\begin{equation}
dI_{\rm{\gamma}}=\rho \beta^{2} \frac{dN_{\rm{\gamma}}}{dE_{\rm{\gamma}}} dV = \beta^{2} \frac{dN_{\rm{\gamma}}}{dE_{\rm{\gamma}}} dm_{\rm{cell}} 
\label{21withbeta}
\end{equation}
In the applications of the present work, by exploiting the advantages of PLUTO, hydrodynamic simulations of $\gamma$-ray emissions from the SS433 MQ system are performed. This description, despite its complexity, is clearly more realistic and advantageous than that of a steady-state model.

\subsection{$\gamma$-ray emission calculations}

The main elementary particle reactions that produce $\rm{\gamma}$-rays through p-p interaction are the pion decay channels

\begin{equation}
pp \rightarrow pp \pi^{0}+\cal{F} \, 
\end{equation}

\begin{equation}
pp \rightarrow pn \pi^{+}+\cal{F} \, .
\end{equation}
($m_{p}=1.67 \times 10^{-24}$ g and $m_{\pi}=2.38 \times 10^{-25}$ g). $\cal{F}$ stands for the typical 'fireball' comprising $\pi^{0}$ and $\pi^{+}\pi^{-}$ channels (Reynoso et al. 2008). Our quantitative description of the above reaction within the jet will mainly proceed (Reynoso et al. 2008, Kelner et al. 2006, Romero et al. 2003, Christiansen et al. 2006) with the calculation of the infinitesimal emissivity $dN_{\rm{\gamma}}/dE_{\rm{\gamma}}$ directly from the hydrodynamical properties of the model jet, as supplied by a check point of the PLUTO code, that constitutes a time instant of a run. For each computational cell, in a 3D mesh, the hydro-quantities of density, pressure, temperature and the three components of the velocity are provided in the form of multi-dimensional arrays. These quantities, along with a number of further assumptions and simplifications about the system, will allow the calculation of the emission coefficients, $\epsilon_{\rm{\nu}}$, at every computational cell of the 3D hydrodynamical model grid. The hydrodynamical jet model, is an improvement over a steady-state one (Romero et al. 2003, Christiansen et al. 2006, Reynoso et al. 2008). The production of synthetic images from the data are performed using the LOS code, where the radio emission and absorption coefficients have been replaced by their $\rm{\gamma}$-ray equivalents, $\epsilon_{\rm{\nu}}$ and $k_{\rm{\nu}}$.

Assuming that the $\rm{\gamma}$-rays in question are created mainly from the collision of fast protons with cold ones, and denoting the energy of the observed $\rm{\gamma}$-ray as $E_{\gamma}$ (measured in \rm{GeV}), the number density of the thermal (bulk flow) jet protons, $n_{\rm{sp}}$, is set equal to the hydrodynamic number density of the PLUTO hydrocode, $n_{\rm{sp}}=n_{\rm j(PLUTO)}$. As mentioned before, the non-thermal proton energy distribution resembles a power-law with its index $\alpha$ set equal to $\alpha$=2 (Rieger et al. 2007, Blandford \& Eichler 1987, Reynoso et al. 2008). In the LOS code, the instantaneous angle between the jet axis and the LOS, for the slowly precessing jets of the SS433 system, is taken for simplicity $i_{j}=1.38$ radians. This angle changes with the jet precession, however for the model checkpoint where we produce a synthetic image of the jet, we keep it fixed, and equal to the above value.

By defining the energy fraction 
\begin{equation}
x=\frac{E_{\gamma}}{E_{\rm{p}}} \, ,
\end{equation}
we take
\begin{equation}
E_{\rm{p}}=\frac{E_{\rm{\gamma}}}{x} \, .
\end{equation}
The observed $\rm{\gamma}$-ray energy, $E_{\rm{\gamma}}$, is kept constant during the composition of an artificial synthetic image, but for every $E_{\rm{\gamma}}$, we get contributions from many different (high-energy) proton energies, each with its own proton population. The above procedure is considered for every computational cell. The intensity of the emission is computed on the basis of the physical properties of the cell that reflect the hydrocode variables at that cell. This way, the emissions are obtained everywhere, for a $\rm{\gamma}$-ray energy value. Relying on the discussion of Kelner et al (2006) and Reynoso et al. (2008) and starting from Eq. \ref{21blackjack}, we arrive at the relations (see Appendix \ref{intapp}):

\begin{equation}
\Delta I_{\rm{\gamma}}=\rho_{\rm{hd}}\frac{dN_{\gamma}}{dE_{\rm{\gamma}}} \Delta V_{\rm{cell}} = m_{\rm{\pi}} n_{\rm{sp}} \beta^{2} \frac{dN_{\gamma}}{dE_{\rm{\gamma}}}(\Delta x_{\rm{cell}})^{3}\,  \, .
\label{digammalastmaintext}
\end{equation}
which are central expressions for our model intensity calculations.


\subsection{Hydrodynamical code employed}

As mentioned before, in the present work the hydrodynamical simulations of the SS433 system are performed using the PLUTO astrophysical code (Mignone et al. 2007) where the jets are approximated as purely hydrodynamical (HD) relativistic flows. In the employed HD approximation, the magnetic field $B$ is dragged with the flow and is taken to be tangled. It, therefore, allows such a coupling of the jet material that makes viable the fluid approximation within the jet (Ferrari 1988, Shu 1991, De Young 2002). On the other hand, the field would not affect the dynamics of the flow (which is possible in the magnetohydrodynamic approximation, or MHD for short). The PLUTO code employs the number density of ejected particles, $n_{\rm{sp}}$, which is dynamically important, whereas the non-thermal proton density, $n_{\rm{fp}}$, even though dynamically unimportant, is nevertheless radiatively significant. 


In order to generally fit the SS433 system (Fabrika 2004), a generic microquasar setup is used,  SS433 was modelled using the 3D RHD version of the PLUTO code in $xyz$ Cartesian form using the piecewise-linear method (van Leer 1979, Collela 1985). In the current model,  however, only one of the twin jets is represented. Even though a counterjet is presumed to exist at the bottom of the $xz$ plane, outside the model space, this is \emph{not} taken to significantly interfere with the model system. The computational grid is homogeneous and the boundary conditions are 'reflective' at the $xz$ plane (base of the jet) and 'outflow' at all other planes of the 'box'. The binary separation is kept as a fixed parameter of the model, pertaining to the stellar wind configuration, and not directly related to the mass estimates of the two stars. The grid spans $120 \times 200$ $\times 120$ (in $10^{10}$cm, for $x$, $y$, $z$, respectively), of equal size ($120 \times 180$ $\times 120$ in high resolution). Two different levels of resolution are used, all covering the aforementioned actual grid dimensions: $300 \times 500 \times 300$ and $480 \times 720 \times 480$, for ($x$, $y$, $z$). The computational cells are therefore nearly cubic, with the jet origin in the middle of the $xz$ plane, at the point (60, 0, 60) $\times$ 10$^{10}$cm. The jet advances while precessing around the (60, $y$, 60) line, (parallel to the $y$ axis) and traverses the (60, 0, 60) point. The precession angle is set to be $\delta$=0.2 radians, slightly less than the reported value of 21 degrees for the SS433 jet (Fabrika 2004), in order to accomodate the jet within the computational domain, allowing in this way the use of finer resolution. 

The companion star is supposed to be centered at the (400, 0, 400) point (Filippova et al. 2006), outside the computational domain,  while a collapsed stellar remnant is situated at the jet base, i.e. at (60, 0, 60). The exact orbital separation in SS433 is not known, so this is an order of magnitude estimate. Both objects are orbiting around their common centre of mass. However, as the jet is travelling at a relativistic velocity u=78000 km/sec=0.26c (Fabrika 2004), the jet crossing time of the computational domain is too small for both stellar objects to move appreciably far from their position during the jet crossing. Therefore, we can safely assume that the stars are immobile for the duration of the simulation, which is of the order of 1000 sec in model time. The stellar orbital speed is about three orders of magnitude less than the jet speed. 

We note that the companion star by itself is not included in the model, we have only its  stellar wind, setup as a density decreasing with distance r away from its centre, proportional to $1/r^{2}$. This approach, of course, makes the simulation less realistic, however the main purpose is to model the jets and their interaction with the winds and interstellar medium. So, for the efficiency of the model, the effects of the jet on the surface and also in the vicinity of the companion star are neglected. Along these lines, a simplified accretion disk wind is included, without the disk itself. For simplicity, the accretion disk wind is not taken to precess with the jet, although in reality, the disk may precess with the jet, causing its wind to deviate from our simple configuration. The jet input radius is fixed at $2 \times 10^{10}$ cm (for technical reasons a constant cell size is used). The jet mass flow is calculated below (Section \ref{jetmass}). 

On the basis of the orbital separation of the two stars, the stellar wind origin does not coincide with the jet base at the scale of the simulation. Furthermore, the jet also travels through a 'halo' produced from the accretion disk, which is centered at the jet base (this matter is commonly called the accretion disk wind). The accretion disk radius is taken to be 2$\times 10^{11}$ cm, centred at the jet base, which is also the location of the compact object. In our simulation, the accretion disk wind is effectively restricted in a cylinder of radius 2$\times$10$^{11}$cm and a density falling off as $1/r^{2}$ away from the accretion disk surface, i.e. along the $y$ axis.
\begin{figure}
\vspace*{7.4cm}
\centering
{\includegraphics{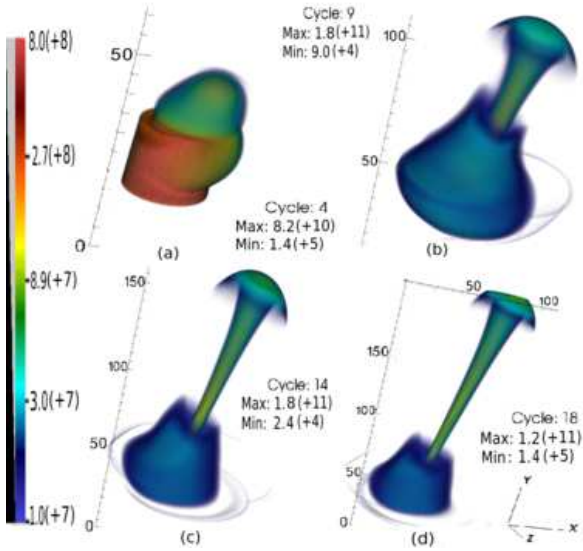}}
\caption{Density evolution over the course of the jet propagation of Scenario A (see Table \ref{Table1}), depicted in pictures (a) to (d). Additional ambient matter piles up ahead of the jet, and to its sides (especially around the jet base), largely originated from the accretion disk wind. The resolution in this simulation is higher ($480 \times 800 \times 480$) than that of scenario B, leading to a more detailed representation of the local dynamical effects.}
\label{run2densitymulti}
\end{figure}

\subsection{LOS code methodology}

The various stages followed in the computational procedure are shown in Appendix \ref{layout} (Fig. \ref{schematicdiagram} is a flow diagram of the simulations). The hydrocode is run and a precessing jet is modelled. At some point in the simulation, the computational space of the model is dumped to a file and the data are then processed with the LOS code, in order to produce a synthetic $\gamma$-ray image of the system. At each point along a LOS, $\rm{\gamma}$-ray emission and absorption coefficients are provided using the formalism of Appendix \ref{intapp}. 

\begin{figure}
\vspace*{7.3cm}
\begin{center}
{\includegraphics{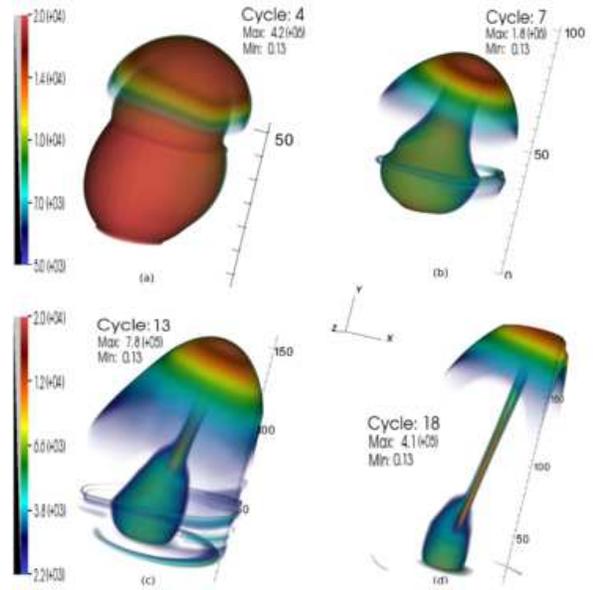}}
\end{center}
\caption{Scenario A: Pressure evolution in and around the jet (run1 parameters in Table \ref{Table1}). The shock front of the jet is expanding, both in the forward direction and sideways. Remnants from the initial accretion disk wind are located around the jet-base, forming a sizeable mass of pressurized material. This concentration appears to persist throughout the simulation, forming a 'halo' around the jet base, as shown in pictures (c),(d).}
\label{run2pressuremulti}
\end{figure}

The LOS code employed in Smponias and Kosmas (2011), originally constructed for radio emission, has been modified in order to be used for $\rm{\gamma}$-rays. To this end, the emission and absorption coefficients are those of Reynoso et al. (2008), suitable for our hydrocode calculations. For neutrinos, only the emission coefficient is applicable (no absorption). The LOS code is run in IDL, while the emission and absorption coefficients are calculated in Mathematica, through a routine that collaborates with the IDL LOS code. A synthetic image is then created that could be directly compared with observations. A spectral energy distribution from the model is also produced, treating the model jet as a point source in the sky.

\begin{figure}
\vspace*{7.7cm}
\begin{center}
{\includegraphics{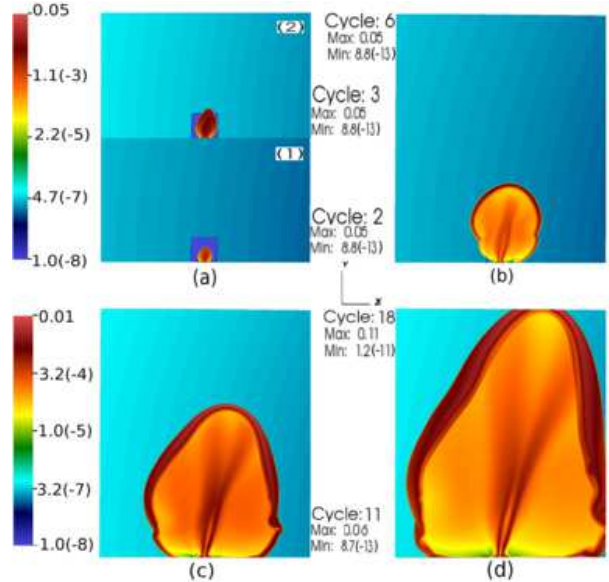}}
\end{center}
\caption{Scenario A: Evolution of temperature in typical 2-dimensional slices, on a plane cut parallel to the jet axis, at a polar angle (azimuthal direction) of 45 degrees. The low gradient of the ambient density leads to more shock front expansion in the forward direction than sideways, as the jet travels away from its base.}
\label{run2temperaturemulti}
\end{figure}

More specifically, the data of a snapshot from the PLUTO hydrocode is transferred, in the form of 3D data arrays (density, velocity vector, pressure: $\rho, u_{\rm{x}}, u_{\rm{y}}, u_{\rm{z}}, P$) to the routine that performs the LOS integration. Then, the LOS code crosses the 3D hydro data volume with multiple parallel LOS's, each of which terminates at a pixel on an observation plane, where a synthetic image is formed. Along each LOS, the equation of radiative transfer (Eq. \ref{1dradtransfer}) is solved. At each model cell position, the local emission coefficient is calculated, using the relevant hydrodynamical properties of the matter and also the radiative properties of the distribution of the high energy protons (no sideways scattering is considered along the LOS). The total intensity, obtained from the summation over each individual LOS is assigned to the intensity of the pixel where the LOS intersects the imaging plane. 

Finally, the synthetic image is normalized as follows: The total model intensity, equal to the sum of individual pixel intensities (excluding some possibly extremely high values, at the edge of the domain near the jet base, which are filtered out with an upper limit function), is set equal to the observed $\rm{\gamma}$-ray luminocity from the system at a given frequency. The resolution of the synthetic images is higher than currently available $\rm{\gamma}$-ray observational resolution from terrestrial Cherenkov arrays (e.g. Orellana et al. 2007a, Hayashi et al. 2009, Saito et al. 2009), and from orbital platforms (e.g. Geldzahler et al. 1984, Abdo et al. 2009). Nevertheless, this margin allows for some exploration of various system configurations, all of which may still match the actual situation.

During the integration along a LOS, the speed at which the LOS is 'drawn' through the 3D hydro data volume is taken to be the speed of light, i.e. much higher than the speed of the jet-flow (u$_{\rm{jet}}$). In this manner, the same data snapshot is used along an individual LOS. If the jet is relativistic, time delay effects must be taken into account and inputs from multiple successive snapshots have to be considered along the same LOS. Because the jet speed of SS433 is intermediate (0.26c), as an approximation we employ the non-relativistic imaging code.

\section{Results and Discussion}
\label{results}

In the present Section, $\rm{\gamma}$-ray emission in jets of the SS433 microquasars are simulated, by using the relativistic hydrocode PLUTO, in a steady-state model and in two hydrodynamic scenarios. The required $\rm{\gamma}$-ray emission coefficients are calculated at every cell. Finally, by integrating along the lines of sight, synthetic images are produced and the total emission is obtained.

\subsection{Steady-state model image}

As the first step of our calculations, we evaluate the $\rm{\gamma}$-ray emission of a steady-state jet. Precession is ignored at this stage, but it is included in the hydrodynamic model (see below). In Figure \ref{losimagehj88}, a synthetic image of a steady state model jet, whose opening angle may decrease along the jet axis (equation \ref{cone-crosec3}, and also Hjellming \& Johnston 1988) is illustrated. In this case, we have set the sideways expansion index of Eq. \ref{Eq-conicmodel} equal to $\ell$=0.5. As can be seen, the intensity decreases along the jet axis, while the jet expands sideways at a slower rate indicating the influence of the surrounding stellar wind.

\begin{figure*}
\vspace{5.0cm}
\begin{center}
{\includegraphics{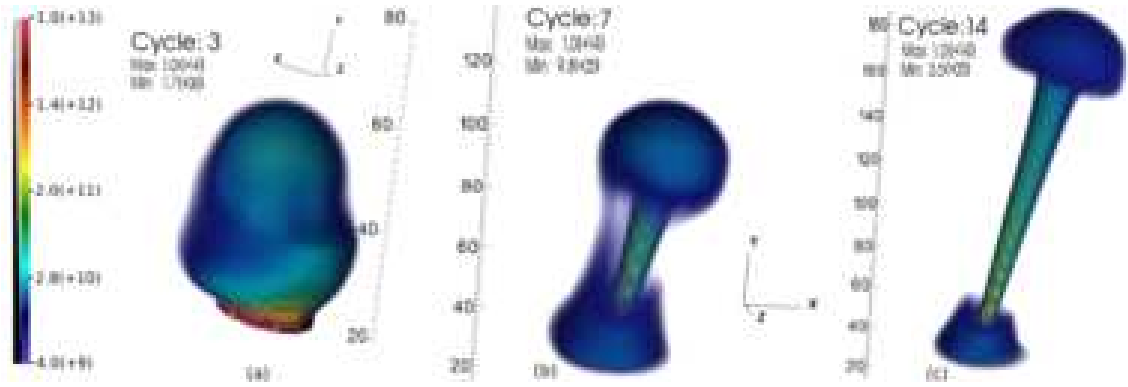}}
\end{center}
\caption{Evolution of the density (logarithmic plots) for the parameters of Scenario B (run2), for a jet quite heavier than both surrounding winds, leading to a faster jet crossing of the model space. The jet mass flow rate is now more akin to the estimates done for SS433.}
\label{run4densitymulti}
\end{figure*}

\begin{figure*}
\vspace{5.4cm}
\begin{center}
{\includegraphics{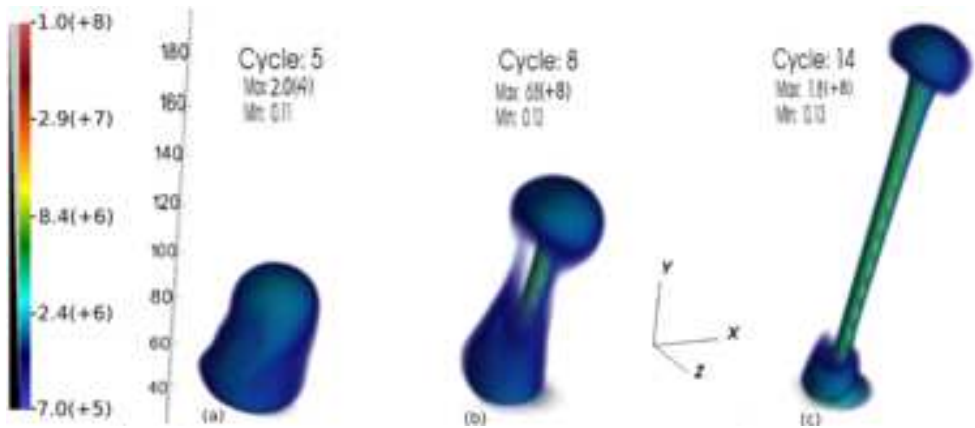}}
\end{center}
\caption{Pressure evolution for the model jet of Scenario B (run2). The higher density of the jet dominates the system dynamics, leading to comparatively poorer features around the jet, as compared to the lighter jet of scenario A (run1). An overpressure appears around the jet base, due to the mass leftover from the disturbed accretion disk wind (run2 parameters listed in Table \ref{Table1}).}
\label{run4pressuremulti}
\end{figure*}

\subsection{Hydrodynamical simulations}\label{hydrosims}

\subsubsection{Our jet model among other scenarios}

Among the great number of possible setups (e.g. Mirabel \& Rodr{\'{\i}}guez 1999, Fabrika 2004, Perez and Blundell 2011) authors have used for the jet and the ambient density, we choose two characteristic cases (scenarios). The corresponding sets of parameters (see Table \ref{Table1}) describes two representative situations of the jet. 

The first scenario (A) features a slowly precessing jet of lower density than that of the stellar wind and accretion disk wind. In the second scenario (B), the jet is quite heavier than both winds, aiming to account for the possibility of a dense jet beam containing the estimated jet mass flow of SS433. The heavier jet of scenario B passes through the winds easier, while interacting rather weakly with them. In both cases, the jet begins to expand into the accretion disk wind, but at a smaller rate, due to the increased density of that wind versus the stellar wind. As soon as the jet head reaches the stellar wind region, the jet's expansion rate greatly increases, especially sideways, in the form of a side shock that accumulates ambient matter. At the same time, the accretion disk matter is expelled outwards from the vicinity of the jet base forming a 'ring' around the jet. The accretion disk wind is swept in a prominent way, being denser than the stellar wind, leading to the creation of a 'halo' around the jet base. The above 'halo' develops throughout the model run, indicating its persistence till the jet reaches its lobe in the W50 nebula and resembling the structure reported in Blundell \& Hirst (2011).

\subsubsection{Description of the runs in the two scenarios}\label{runs}

\paragraph{Scenario A, light jet:}

In this scenario (run1), the accretion disk wind and the stellar wind densities (Table \ref{Table1}, run1 column) cause the jet to advance slower than in run2, with more material accumulating behind the jet-head and within the jet cocoon. The simultaneous evolution of the density and pressure, as the jet propagates, is illustrated in Figs. \ref{run2densitymulti} and \ref{run2pressuremulti}, respectively. As a result, an active jet cocoon appears as shown in Fig. \ref{run2temperaturemulti}. The jet initially crosses the accretion disk wind, which is denser than the stellar wind, before moving into the stellar wind itself, where it advances faster due to smaller resistance. The jet's sideshock now moves closer to the inner stellar wind of the companion star, causing increased activity at the side of the jet towards the location of the star (Fig. \ref{run2temperaturemulti}).
\paragraph{Scenario B, Heavy jet:}

\begin{figure}
\vspace{8.0cm}
\begin{center}
{\includegraphics{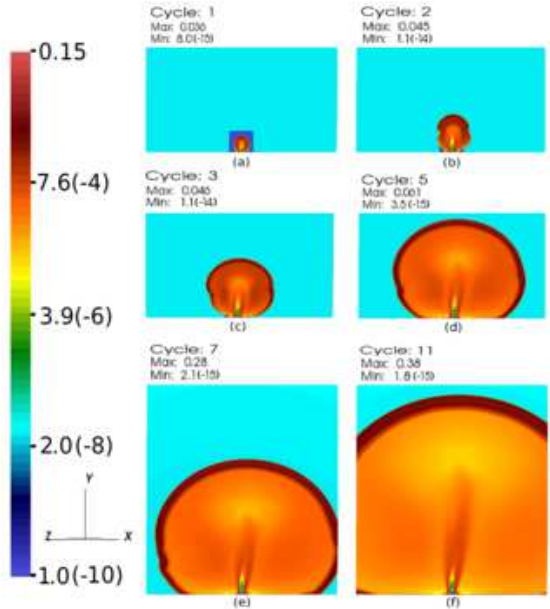}}
\end{center}
\caption{2-dimensional plots of the jet temperature of Scenario B (see run2 parameters in Table \ref{Table1}), as in Fig. \ref{run2temperaturemulti}. The -now faster- advance of the jet is characterized by an expanding shock front. Expansion is initially slower, till the jet crosses the disk wind. Upon entry to the stellar wind region, a higher rate of expansion occurs, especially to the sides.}
\label{run4temperaturemulti}
\end{figure}

In the second scenario, the jet is assumed to be heavier than its surrounding winds, which in turn are also somewhat heavier (larger density) than those of scenario A, leading to a faster crossing of the computational domain (Fig. \ref{run4densitymulti}). Sideways expansion is rapid yet limited spatially as the increased jet-mass density allows for a faster, almost ballistic, 'sweep' of the wind matter. Leftovers from the displaced matter of the accretion disk wind can be seen piling up, with increased pressure, around the jet base. The jet forms a funnel that transfers mass outwards at an increased flow rate. The properties of the jet's surroundings are now less pronounced, as the expansion meets with reduced resistance from ambient matter (Fig. \ref{run4pressuremulti}). The dynamical behaviour of the jet dominates the hydro simulation, with the wind's matter giving way to the jet (Fig. \ref{run4temperaturemulti}). This case covers the possibility of a very dense jet inflow at the source, thus enriching nearby interstellar matter with signifcant mass outflow. 

\subsubsection{Model jet mass output}
\label{jetmass}

For the heavy jet (run 2), we focused on the reproducibility of the total jet emission, as compared to estimates of the jet kinetic "luminocity" $L_{k}=10^{39}$ergs$^{-1}$ (e.g. Fabrika 2004). For this model jet, we obtain a kinetic power of $L_{k}=10^{40}$ergs$^{-1}$, leading to a total emitted power of L=10$^{36}$ergs$^{-1}$, for q$_{\rm{rel}}$=0.0001 (see Appendix \ref{q-factor}). The latter result can then be used to determine the normalization constant K$_{1}$ (see Eq. \ref{Jfp-2}), by equating the unnormalized emitted power with the q$_{\rm{rel}}$ fraction of the model's kinetic energy. 

For the light jet scenario A, the density, and correspondingly the kinetic luminocity, are three orders of magnitude smaller. This scenario describes an episodic lower density ejection where the kinetic power of the system is lower. Even though this outburst emits weaker, due to slower advancement of the jet front, the emission may persist for longer in the inner jet area. As a result, a richer inner jet dynamical environment appears, contributing to a weaker, but more persistent and spatially extended, emission.

\subsection{Results of $\gamma$-ray emission with the radiative code}\label{radcoderesults}

\subsubsection{Line of sight calculation}
\label{generalloscoderesults}

Initially, artificial $\gamma$-ray images were formed using at each model jet location (computational cell), either $\rho$ or $\rho \beta^{2}$ as the emission coefficient. As a first approximation, the absorption coefficient was considered proportional to the local matter density $\rho$, due to the interaction of $\gamma$-ray photons with the jet thermal protons (Reynoso et al. 2008). For technical reasons, the imaging process was presented at less resolution than that of the original hydrocode runs, namely it was decreased to $60 \times 100 \times 60$, by employing PLUTO's attached IDL routine (pload) to regrid the hydro data.

The jet travels across the computational domain decreasing in density as it flows outwards. In general, the emission/absorption coefficients may depend on the local hydrodynamic quantities. However, the energy integrals, O1 and O2 (see equations \ref{O1} and \ref{O2} in Appendix \ref{intapp}), that provide the emission coefficients, do \textit{not} depend on hydro quantities for the $\gamma$-ray studies, and the radiative calculation can be separated from the hydro quantities. Since the applied types of calculation are interconnected, transfer of 3D data back and forth between the IDL and Mathematica codes is required. Nevertheless, separate calculations can be performed in our case, helping shorten the execution time of the runs, since the shape of the high energy proton's distribution $n_{\rm{fp}}$ is independent of the velocity $u$ and density $\rho$ they may  be taken out of the radiative integrals O1 and O2.

In practice, the LOS integration is performed using as the emission coefficient first $\rho$ and then $\rho \beta^{2}$ while the absorption coefficient is kept proportional to $\rho$. The synthetic image is then formed, and subsequently the values of all of its pixels are added up to provide the total intensity released from the object. The $\gamma$-ray emission calculation is treated with a Mathematica code (for a unit proton number density). Practically, for each of the two scenarios (runs), the density and velocity output data originating from a checkpoint of the PLUTO hydrodynamic runs, were saved into disk. Finally, the 3-D data grid were imported to IDL, where they were 'crossed' by many parallel lines of sight (LOS). Along each LOS, the radiative transfer equation \ref{1dradtransfer} (without scattering) was solved, using our radiative transfer code. 

\subsubsection{Simulated $\gamma$-ray emission and absorption}
\label{mathematica}

In addition to the emission obtained from Eq. \ref{digammalast} at each computational cell (represented by either $\rho$ or $\rho \beta^{2}$), the $\gamma$-ray absorption coefficient for a cell, k$_{\nu}$, is computed via the relation 
\begin{equation}
k_{\nu}=\rho \left( \frac{\sigma_{\gamma N}}{m_{p}} \right) 
\end{equation}
where $\sigma_{\gamma N}$ is the cross section for the combined effect of photo-pion and photo-proton pair production processes, i.e. $\sigma_{\gamma N}$=$\sigma^{(\pi)}_{p \gamma}+\sigma^{(e)}_{p \gamma}$ (Atoyan \& Dermer 2003, Begelman et al. 1990, Reynoso et al. 2008). The simulated emission of $\gamma$-rays is performed using the same emission calculation throughout the computational domain. Although the hydrodynamic variables differ from location to location, they can be separated and taken out of the emission integrals (see Appendix \ref{intapp}). The LOS artificial images are subsequently drawn using the hydro data and are multiplied by the emission integrals in order to obtain the final images. 

\begin{figure}
\begin{center}
\vspace{6.0 cm}
{\includegraphics{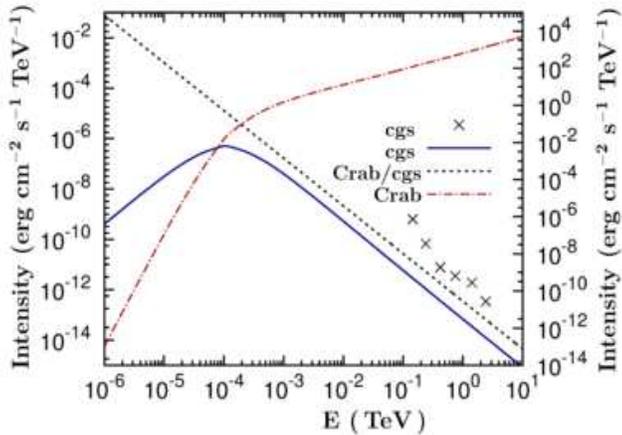}}
\caption{\emph{Right} vertical scale: The purple line (short dashes, in cgs units) represents the Crab spectrum (note that the right vertical scale, used for the Crab Spectrum, is different than the left one). Left vertical scale: Spectral energy distribution of the $\gamma$-ray emission, for unit hydrodynamical density (red solid line, in cgs units). The total emission (red solid line) may be obtained (within a normalization constant) by adding up, at every frequency, the emission from all pixels of an artificial image. The blue line (long dashes, \emph{Crab units}) is the SED expressed in Crab units (left vertical scale, but now measured in Crab units). Observational upper limits from MAGIC (Saito et al. 2009) are also shown as crosses (in cgs units, left vertical scale).
 } 
\label{figuresed}
\end{center}
\end{figure}

In calculating the emission coefficients, many different values for the energy of the $\gamma$-rays are used from 1 \rm{MeV} to 5 \rm{TeV} (Fig. \ref{figuresed}). For each case, the energy integrals (O1, O2) vary as functions of $E_{\rm{\gamma}}$ only (there is no spatial dependence). We remind that for neutrino emission, there exists dependence on spatial parameters of the integrals in question (Reynoso et al. 2008). The resulting artificial images are, then, multiplied by the results of the emission coefficient calculation, performed in Mathematica.

\begin{figure}
\begin{center}
\vspace{4.2 cm}
{\includegraphics{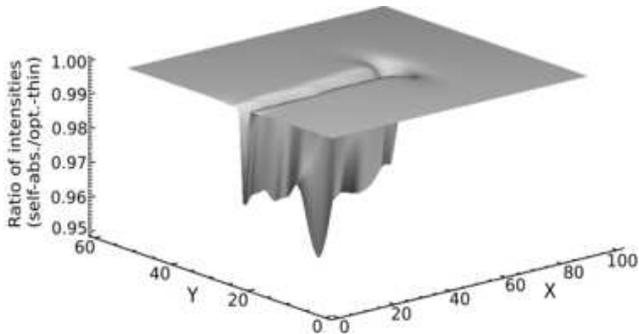}}
\caption{Absorption of $\gamma$-rays from the jet matter, at $E_{\gamma}$=200 GeV. The ratio of intensities with/without absorption, of a synthetic $\gamma$-ray image, is plotted for the heavy jet Scenario B, displaying a self-absorption of up to 5 percent. The absorption peaks along the main jet body. 
 } 
\label{figureabs}
\end{center}
\end{figure}

Concerning absorption, calculations are performed in this paper for only a representative $\gamma$-ray energy, because more demanding computations are required, and then less approximations are valid (for further analysis see  Smponias \& Kosmas 2013, Smponias \& Kosmas, in prep.). At $E_{\rm{\gamma}}$=200 GeV, a rough estimate of absorption, is obtained as follows: If $\rho / m_{\rm{p}}$=10$^{14}$gcm$^{-3}$, $\sigma_{\gamma \rm{N}}$=2$\times$10$^{-26}$cm$^{-2}$ and dl=5$\times$10$^{10}$cm, the ratio of the total intensities at $E_{\rm{\gamma}}$=200 GeV, obtained with and without absorption respectively, is at maximum around 5$\%$, depending on the hydrocode snapshot used (Fig. \ref{figureabs}). As mentioned before, here only absorption from matter is considered while absorption from X-rays will be performed elsewhere (Smponias \& Kosmas, in prep.).

The LOS integration relying on hydrodynamic data provides the emitted $\gamma$-ray power per unit energy interval from the model system. Then, the intensity can be approximated (Doppler boosting is ignored, since the angle of observation is 79 degrees from the jet axis and the jet speed is 0.26c see e.g. De Young 2002) by considering it as -homogeneously- spread across the surface of the fiducial sphere, centred on the microquasar and having a radius equal to the distance to the observer (in our case the Earth). In order to be able to match the calculated values to the observed ones, a normalization is required (see below). The resolution employed in synthetic $\gamma$-ray imaging is higher than currently available observational resolution. Nevertheless, the model may fit improved future observations as soon as they would become available. The sum of model intensities coming from all LOS's constitutes the model source intensity, at a given $\gamma$-ray energy, $E_{\gamma}$. This quantity is calculated over a range of $E_{\gamma}$ leading to a spectral energy distribution (SED) of the model (Fig. \ref{figuresed}).

Nevertheless, the quantities of density and velocity have been taken out of the energy integrals O1 and O2, and consequently, a single SED plot suffices to cover all possible cases of the current model. We note that, the total intensity is calculated from the LOS code (spatial integration) using local density and velocity values. Subsequently, this sum multiplies the SED plot, while normalization is performed in relation to the jet's energetics. The spatial detail inherent to the model, due to the hydrocode and LOS code used, is not exploited here. However, the hydrodynamic approach is still more realistic than the steady-state one, offering the potential to explore the dynamical evolution of the system, especially in episodic jet ejections.

Before closing our discussion on the present simulations, we note that, for rather low energy $\gamma$-rays, space-born platforms, such as \emph{Fermi} (NASA orbital telescope) and \emph{INTEGRAL} (ESA satellite), offer an important relevant body of observations. Very high energy $\gamma$-rays, which constitute the larger portion of the spectrum modelled in the present paper (generally above the threshold of 30 \rm{GeV}), can also be studied with ground based Cherenkov telescopes such as HESS, MAGIC and VERITAS, as well as the upcoming CTA. In view of the availability of such data, hydrodynamical jet models, combined with artificial imaging, provide more realistic estimates of the physical conditions within the jet and surrounding environments, than a steady-state model. Such modelling attempts may offer insight to open questions regarding the $\gamma$-ray jet emission from microquasars.

Specifically, the Cherenkov Telescope Array (CTA) is expected to offer increased sensitivity for $\gamma$-ray detection from potential MQ sources (Actis et al. 2011). Using the full CTA resolution as a reference (even though SS433 lies in the north hemisphere), one may compare to our model predictions based on energetic arguments. For a jet kinetic 'luminocity' of $L_{k}$=10$^{40}$ ergs/s, and a relativistic proton power factor of 10$^{-4}$, we obtain an emitted power $L_{\gamma}$=10$^{36}$ erg/s. The latter should then be divided by the surface of a sphere centred at SS433 and having a radius equal to the distance to Earth, S=4$\pi R^{2}$, leading to $L_{\gamma}$/S=2$\times$10$^{-10}$ ergs$^{-1}$cm$^{-2}$. 

Finally, by setting the area under the SED plot of the model data (Fig. \ref{figuresed}) equal to the aforementioned integrated flux density at Earth (the distance to Earth has been already taken into account), we obtain the calibration constant K$_{1}$=2$\times$10$^{7}$. Using the latter constant value, we  plot the model differential flux density, in Crab units (Fig. \ref{figuresed}), and compare it to the CTA estimate (Actis et al. 2011, figs. 19 \& 20). Our result is at most an order of magnitude less than the minimum projected full CTA sensitivity. On the other hand, the light jet gives three orders of magnitude lower total emission, therefore its detection lies beyond present day capabilities. Saito et al. (2009) using the MAGIC detector array, provided upper limits for the SS433 $\rm{\gamma}$-ray emission for a range of observing energies, from 100 \rm{GeV} up to the TeV range. These exceed the model values at their corresponding energies (Fig \ref{figuresed}).

\section{Summary and Conclusions}

At first, in order to test the methods employed, a steady state model was reproduced, using computational tools (LOS code, $\gamma$-ray emission code). $\gamma$-ray emission along a conical jet was then calculated. A precessing jet was then modelled using the relativistic hydrodynamical code (PLUTO). Furthermore, the results were processed using the LOS code (assuming that the flow velocity $u$ is much smaller than the speed of light, u$_{\rm{jet}}$=0.26c), which integrates, ignoring scattering, along lines of sight the equation of radiative transfer. The emission coefficients are, in general, functions of hydrodynamical and radiative parameters.
 
Two representative scenarios were used, in order to study different sets of parameters of the model jet. The intensity resulting from each LOS is assigned to the pixel where the LOS meets the imaging plane of 'observation', forming a synthetic $\gamma$-ray image. Even though in this paper, we mainly focused on $\gamma$-ray emission, absorption and neutrino emission may also be incorporated. The connection between the emission properties and the underlying system dynamics offers useful constrains on a variety of system parameters, such as the jet kinetic luminosity $L_{k}$. The above may be helpful in light of potential future observations, originating from orbital $\gamma$-ray telescopes, from terrestrial Cherenkov detector arrays (e.g. the CTA) and even from underground neutrino detectors. 

As a next step, an RMHD precessing jet can be modelled, allowing the magnetic field $B$ to vary dynamically within the system. Furthermore, the high-energy neutrino emission from such sources can also be reliably simulated.
  
\section{Acknowledgements}
We would like to thank the personnel of the Grid node of the University of Ioannina for their helpful assistance. TS thanks D. Papoulias for assistance with the Mathematica package.
 



\indent
\bsp

\appendix
\section{Estimation of the power fraction q}
\label{q-factor}

 In order to estimate the fraction $q_{\rm{rel}}$ of the total jet power carried by the high energy proton distribution, at every computational cell (Reynoso et al. 2008, equation 4), we first note that
\begin{equation}
q_{\rm{rel}}L_{\rm{k}}=\pi R_{0}^{2} \int\limits_{E'_{\rm{min}}}^{E'_{\rm{max}}} J' E'_{\rm{p}} dE'_{\rm{p}}
\label{eqapp1}
\end{equation}
where J$^{'}$ is the non-thermal proton flux in the jet frame, $L_{k}$ is the jet kinetic luminosity, $R_{0}$ is the presumed jet radius at the jet location where $q_{rel}$ is calculated, and $E'_{p}$ is the energy (in the jet frame) of the high energy protons. Finally, $E'_{\rm{min}}$ and $E'_{\rm{max}}$ represent the minimum and maximum energies of the high energy protons, respectively. By inserting Eqs. \ref{jprimefp1} \& \ref{Jfp-2} in equation \ref{eqapp1}, we get
\begin{equation}
q_{\rm{rel}}L_{\rm{k}}=\pi R_{0}^{2} \int\limits_{{E'}_{\rm{min}}}^{{E'}_{\rm{max}}}\frac{c}{4\pi}(K_{1}n_{\rm{j}} 
\beta^{2}) (E^{'}_{\rm{p}})^{-\alpha} (E^{'}_{\rm{p}}) dE'_{\rm{p}}
\end{equation}
\begin{equation}
=R_{0}^{2} \frac{c}{4} \left( K_{1}n_{\rm{j}} \beta^{2} \right) \int\limits_{{E'}_{\rm{min}}}^{{E'}_{\rm{max}}} 
({E'}_{\rm{p}})^{1-\alpha}dE'_{\rm{p}}
\end{equation}
By putting $\alpha=2$ (Rieger et al. 2007), the latter equation leads to
\begin{equation}
q_{\rm{rel}} L_{\rm{k}} = R_0^2 \frac{c}{4} (K_1 n_{\rm{j}}\beta^2) \ln 
\left( \frac{E'_{\rm max}} {{E'_{\rm min}} } \right) 
\end{equation}
which corresponds to equation 5 of Reynoso et al. (2008).

\label{lastpage}

\section{Schematic representation of the simulation layout}
\label{layout}

\begin{figure}
\begin{center}
\vspace{4.5cm}
{\includegraphics{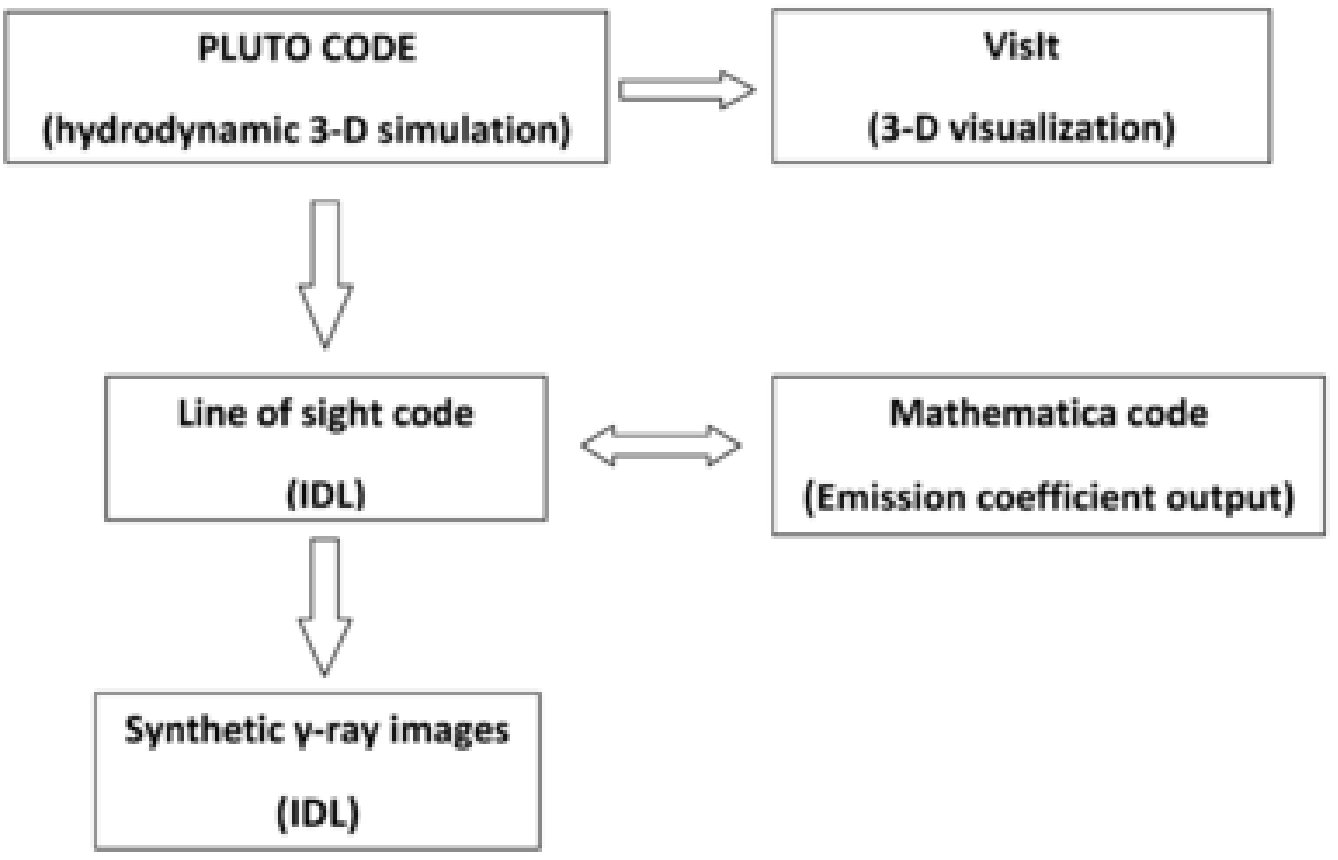}}
\caption{Schematic representation of the simulation layout.}
\label{schematicdiagram}
\end{center}
\end{figure}

A more detailed description, of the steps followed in the dynamic simulations, is illustrated in Fig. \ref{schematicdiagram}. 

\section{Intensity calculation}
\label{intapp}

In this appendix we derive the expression giving the model jet's differential emissivity I. Initially, following Kelner et al. (2006), we define the quantities
\begin{equation}
L=log \left( \frac{E_{\rm{p}}}{1 {\rm TeV} } \right) =log \left( \frac{E_{\gamma}}{1 {\rm TeV} \times x} \right) \, ,
\end{equation}
\begin{equation}
B_{\gamma}=1.3+0.14L+0.011L^{2}
\end{equation}
\begin{equation}
\beta_{\gamma}=\frac{1}{1.79+0.11L+0.008L^{2}}
\end{equation}
and
\begin{equation}
k_{\gamma}=\frac{1}{0.801+0.049L+0.014L^{2}}
\end{equation}
Then, the cross section for inelastic $pp$ scattering reads
\begin{equation}
\sigma^{\rm{pp}}_{\rm{inel}}=(34.3+1.88L+0.25L^{2})
\left[1 - \left( \frac{E_{\rm{th}}}{E_{\rm{p}}} \right)^{4}\right]^{2}
\end{equation}
(in $10^{-27}$cm$^{2}$) where the threshold energy $E_{\rm{th}}$, for production of $\pi^{0}$ mesons, in inelastic $pp$ interactions, is given by (in units of \rm{GeV})
\begin{equation}
E_{\rm{th}}=m_{\rm{p}}+2m_{\pi}+\frac{m_{\pi}^{2}}{2m_{\rm{p}}}=1.22 
\end{equation}

We saw in Section \ref{hydrodynamicaljet} that:
\begin{equation}
J_{\rm{fp}} \equiv J_{\rm{{p}}}= \left( \frac{c}{4\pi} \right) K_{1} \beta_{\rm{sp}}^{2} n_{\rm{sp}}  S(E_{\rm{p}})
\label{jfpjp}
\end{equation}
Also
\begin{equation}
E^{\rm{max}}_{\rm{p}}=3.4\times10^{6} {\rm GeV}
\end{equation}
(Christiansen et al. 2006, Reynoso et al. 2008).
For the production rate $q_{\pi}$, of $\pi^{0}$-mesons, we need the maximum (Reynoso et al. 2008) and minimum energy values as (Kelner et al. 2006) 
\begin{equation}
E_{\pi}^{\rm{max}}=K_{\pi}(E^{\rm{max}}_{\rm{p}}-m_{\rm{p}}c^{2})
\end{equation}
and
\begin{equation}
E_{\pi}^{\rm{min}}=E_{\gamma}+\frac{m_{\pi}^{2}c^{4}}{4E_{\gamma}}
\end{equation}
where $K_{\pi}=0.17$ is the mean fraction of the kinetic energy $E_{\rm{kin}}=E_{p}-m_{p} c^{2}$ transferred, per collision, to a secondary meson (Kelner et al. 2006). We furthermore define the ratios 
\begin{equation}
x_{\rm{max}}=\frac{E^{\rm{max}}_{\gamma}}{E_{p}}
\end{equation}
\begin{equation}
x_{\rm{min}}=\frac{E^{\rm{min}}_{\gamma}}{E_{p}} \, ,
\end{equation}
The latter definitions imply that $x_{\rm{max}}<1$. By replacing E$_{p}$ with $E_{\gamma}/x$ in equation (\ref{jfpjp})  we obtain (Kelner et al. 2006, Reynoso et al. 2008)
\begin{equation}
J_{\rm{p}}=\frac{c}{4\pi} K_{1} \beta^{2} n_{\rm{sp}} S \left( \frac{E_{\gamma}}{x} \right)
\end{equation}
The production rate of $\pi^{0}$-mesons, $q_\pi$, is then defined in terms of $J_{\rm{p}}$
\begin{equation}
q_\pi = \left( \frac{n^{\prime}}{K_\pi}\right) \sigma^{\rm{inel}}_{\rm{pp}}(m_{\rm{p}} c^2 +\frac{E_\pi}{K_\pi}) J_{\rm{p}}
\end{equation}
which, by using equation (\ref{jfpjp}), takes the form
\begin{eqnarray}
q_{\pi} = \left(\frac{n^{\prime}}{K_{\pi}}\right) \sigma^{\rm{inel}}_{\rm{pp}} \left( m_{\rm{{p}}}c^{2}+\frac{E_{\pi}}{K_{\pi}} \right) \nonumber \times \\ 
 \left( \frac{c}{4\pi} \right) K_{1} n_{\rm{sp}} \beta^{2} S \left( m_{\rm{p}}c^{2}+\frac{E_{\pi}}{K_{\pi}} \right)
\label{qpi}
\end{eqnarray}
There are two possibilities for the gamma ray emissivity, $dN_{\gamma}/dE_{\gamma}$, depending on whether $E_{\gamma}$ lies above or below the value of 100 \rm{GeV}. We assume (Kelner et al. 2006) that 

\begin{equation}
\frac{dN_{\gamma}}{dE_{\gamma}} = 
\begin{cases} 
O_{1} &\mbox{, for } E_{\gamma}> 100 GeV \\
O_{2} &\mbox{, for } E_{\gamma}< 100 GeV 
\end{cases}  
\end{equation}
%
where
\begin{equation}
O_{1}=\int_{x_{\rm{min}}}^{x_{\rm{max}}} \frac{dx}{x} \, \sigma^{\rm{inel}}_{\rm{pp}}(E_\gamma/x) J_{p}F_{\gamma}
\label{O1}
\end{equation}
\begin{eqnarray}
O_2 =2 \int_{E_{\pi}^{\rm{min}}}^{E_{\pi}^{\rm{max}}} d E_\pi \frac{q_{\pi}}{\sqrt{E^{2}_{\pi}-m^{2}_{\pi} c^{4}}}\, .
\label{O2first}
\end{eqnarray}
Using equation (\ref{qpi}) the latter integral becomes 
\begin{eqnarray}
O_2 =
2 n^{\prime} \int_{E_{\pi}^{\rm{min}}}^{E_{\pi}^{\rm{max}}} d E_\pi 
F(m_p c^2 + \frac{E_\pi}{K_\pi}) \sigma^{\rm{inel}}_{\rm{pp}}
(m_p c^2 + \frac{E_\pi}{K_\pi}) \times \nonumber \\ 
n_{\rm{sp}} \beta^{2} \frac{K_1}{K_\pi} 
\left(\frac{c}{4\pi} \right) \times 
\left( {E_\pi^2 - m_\pi^2 c^4}\right)^{-1/2}
\label{O2}
\end{eqnarray}

Where the function $F_{\gamma}$ is written as (Kelner et al. 2006):
\begin{eqnarray}
F_{\gamma}=B_{\gamma} \frac{log(x)}{x} \left(\frac{(1-x^{\beta_{\gamma}})}{1+k_{\gamma}x^{\beta_{\gamma}}(1-x^{\beta_{\gamma}})} \right)^{4} \nonumber \\ 
\times \left( \frac{1}{log(x)}-\frac{4\beta_{\gamma}x^{\beta_{\gamma}}}{1-x^{\beta_{\gamma}}}-\frac{4k_{\gamma}\beta_{\gamma}x^{\beta_{\gamma}}(1-2x^{\beta_{\gamma}})}{1+k_{\gamma}x^{\beta_{\gamma}}(1-x^{\beta_{\gamma}})} \right)
\label{fkelner}
\end{eqnarray}
An alternative simpler expression for $F_{\gamma}$ has been suggested, by Hillas (2005) (see also Kelner et al. 2006)
\begin{equation}
F_{\gamma \rm{H}}=3.06 \exp(-9.47 x^{3/4})
\label{fhillas}
\end{equation}

From their definition, it becomes obvious that O1 and O2 should match at $E_{\gamma}$= 100 \rm{GeV}, so the constant $n'$ in $O_{2}$ could be adjusted accordingly. In both O1 and O2, the quantity $J_{p}$ is proportional to the hydrodynamic quantities of jet density and jet velocity. Therefore, the calculation of the emission coefficient can be simply performed just once for the whole grid (instead of once per each computational cell, as is the usual case) and the emission obtained then multiplies the hydro-based result of all cells.

From Eq. \ref{21withbeta}, the differential intensity per unit volume (computational cell volume) has been written as 
\begin{equation}
\frac{dI_{\gamma}}{dV} = \rho_{\rm{hd}} \beta^{2} \frac{dN_{\gamma}}{dE_{\gamma}} = m_{\rm{p}} n_{\rm{sp}} \beta^{2} \frac{dN_{\gamma}}{dE_{\gamma}}
\end{equation}
($\beta^{2}$ accounts for the emission being stronger wherever matter moves at relativistic velocities). Furthermore, assuming that for a cell
\begin{equation}
dV \simeq \Delta V_{\rm{cell}}=(\Delta x_{\rm{cell}})^{3}\, cm^{3} \, .
\end{equation}
we write
\begin{equation}
\Delta I_{\gamma}=\rho_{\rm{hd}}\frac{dN_{\gamma}}{dE_{\gamma}} \Delta V_{\rm{cell}} = m_{\rm{p}} n_{\rm{sp}} \beta^{2} \frac{dN_{\gamma}}{dE_{\gamma}}(\Delta x_{\rm{cell}})^{3}\, \, .
\label{digammalast}
\end{equation}
This equation is used for calculating the model emission in Section \ref{results}. Similarly, for the non-relativistic case the $\beta^{2}$ coefficient may be omitted.
\end{document}